\documentclass[twocolumn,floatfix]{revtex4}

\usepackage{dsfont} 
\usepackage{bm} 

\usepackage{bbold} 
\usepackage{xcolor}
\usepackage{times}
\usepackage{physics}
\usepackage{graphicx,braket}
\usepackage{hyperref}
\hypersetup{colorlinks=true,linkcolor=blue,citecolor=blue}
\usepackage{amsmath,amssymb,amsfonts}
\usepackage{wrapfig}
\usepackage{siunitx}

\newcommand{\gdos}{\mathinner{g^{(2)}(0)}}
\newcommand{\mean}[1]{\langle #1 \rangle}

\usepackage{cleveref}
\crefname{equation}{Eqs.}{Eqs.}
\Crefname{equation}{Equation}{Equations}
\crefrangelabelformat{equation}{(#3#1#4--#5#2#6)}
\crefmultiformat{equation}{Eqs. (#2#1#3}{, #2#1#3)}{#2#1#3}{#2#1#3}
\Crefmultiformat{equation}{Equations (#2#1#3}{, #2#1#3)}{#2#1#3}{#2#1#3}

\begin{document}

\flushbottom
\title{Photon correlation spectroscopy as a witness for quantum coherence}

\author{Carlos S\'anchez Mu\~noz }
\author{Frank Schlawin}
\affiliation{Clarendon Laboratory, University of Oxford, Parks Road, Oxford OX1 3PU, United Kingdom}

\newcommand{\down}{\op{g}{e}}
\newcommand{\up}{\op{e}{g}}
\newcommand{\downd}{\op{+}{-}} 
\newcommand{\upd}{\op{+}{-}}
\newcommand{\app}{a^\dagger}
\newcommand{\ssp}{\sigma^\dagger}
\newcommand*{\Resize}[2]{\resizebox{#1}{!}{$#2$}}%

\begin{abstract}
The development of spectroscopic techniques able to detect and verify quantum coherence is a goal of increasing importance given the rapid progress of new quantum technologies, the advances in the field of quantum thermodynamics, and the emergence of new questions in chemistry and biology regarding the possible relevance of quantum coherence in biochemical processes. Ideally, these tools should be able to detect and verify the presence of quantum coherence in both the transient dynamics and the steady state of driven-dissipative systems, such as light-harvesting complexes driven by thermal photons in natural conditions. This requirement poses a challenge for standard laser spectroscopy methods. Here, we propose photon correlation measurements as a new tool to analyse quantum dynamics in molecular aggregates in driven-dissipative situations. 
We show that the photon correlation statistics on the light emitted by a molecular dimer model can signal the presence of coherent dynamics. Deviations from the counting statistics of independent emitters constitute a direct fingerprint of quantum coherence in the steady state. 
Furthermore, the analysis of frequency resolved photon correlations can signal the presence of coherent dynamics even in the absence of steady state coherence, providing direct spectroscopic access to the much sought-after site energies in molecular aggregates.
\end{abstract}
\date{\today} \maketitle

\emph{Introduction---}The emerging field of quantum thermodynamics assesses the role of quantum fluctuations on thermodynamic properties of meso- or nanoscopic systems~\cite{Sai16, Binder18, Streltsov17, Lostaglio15}. Of particular importance is the possible role of quantum coherence to enhance the efficiency of thermodynamic processes in the quantum regime and the possible gain for technological applications. It has stimulated a lot of theoretical investigations into the detection of coherence~\cite{li12a,marcus19a},  the circumstances when this advantage can be gained~\cite{Scully11, Mitchison15,wertnik18a,wang19a} and how much work can be gained from a quantum system~\cite{Allahverdyan04, Klatzow19, Dorfman18b}. 
In parallel, recent years have also seen an intense debate in the field of chemical physics. Sparked by the discovery of quantum-mechanical coherence in photosynthetic complexes, it was hypothesised that coherent dynamics following the absorption of sunlight might be relevant for the functionality of these complexes, and even form a key ingredient for the high quantum efficiency of solar light harvesting~\cite{Scholes11, Romero17, Scholes17}. 

However, as these discoveries are based on phase-coherent ultrafast laser experiments~\cite{Shaul_book}, it was argued that transient coherences should be irrelevant for natural photosynthesis, and they rather constitute an artefact of the laser usage~\cite{Brumer12, Duan17}. Accordingly, natural photosynthesis instead takes place in a nonequilibrium steady state, and only coherence in such steady state regimes could reasonably be expected to have functional relevance. A growing number of theoretical studies discusses the impact of light statistics on the ensuing dynamics~\cite{Chenu16, Chan18, Dodin19, Fujihashi19}. In addition, a considerable number of theoretical models have been put forth, where steady state coherence is induced by environments~\cite{Scully11, ilessmith14a, Tscherbul15, ilessmith16a,Dodin16, NJP18, Tscherbul18, Brumer18}, both relating to photosynthesis and chemical reactions more generally. Yet to date, there exists no direct spectroscopic probe of these nonequilibrium steady states. The detection of transient coherence in ultrafast laser spectroscopy, too, relies on a comparison to model calculations, in order to infer the presence or absence of coherence~\cite{Romero14, Fuller14}.

In this paper, we introduce photon correlation spectroscopy as a possible tool to measure coherent dynamics and the presence of steady state coherence between excitons in single molecule spectroscopy. Our proposal relies on measurements of photon correlations of the light emitted by single molecular aggregates~\cite{hettich02a,Kruger10,Hildner13, Wientjes14, Liebel18, Kyeyune19} in natural, non-equilibrium situations. In particular, we show how the presence of steady-state coherence can be detected in two-photon coincidence measurements by varying the detection polarization. We further prove that frequency-resolved counting statistics~\cite{delvalle12a,gonzaleztudela13a,sanchezmunoz14b,ulhaq12a,peiris15a,silva16a,Holdaway18} 
can reveal similar information on transient coherent dynamics as two-dimensional spectroscopy~\cite{Cho08, HammZanni}. The analysis of these signals could provide access to the local on-site energies in molecular aggregates, which form one of the greatest unknown in theoretical models of such complexes~\cite{Adolphs07, Cheng09, Milder10}.
Crucially, this information is obtained in a non-equilibrium stationary state, without the need of  phase-coherent ultrafast pulses. 
As photon correlation spectroscopy considers properties of the fluorescence emitted from a sample, it can be carried out with excitation by incoherent light sources, and thus provide insights into nonequilibrium dynamics in conditions of natural illumination of the samples.%

\begin{figure*}[t]
\begin{center}
\includegraphics[width=1\textwidth]{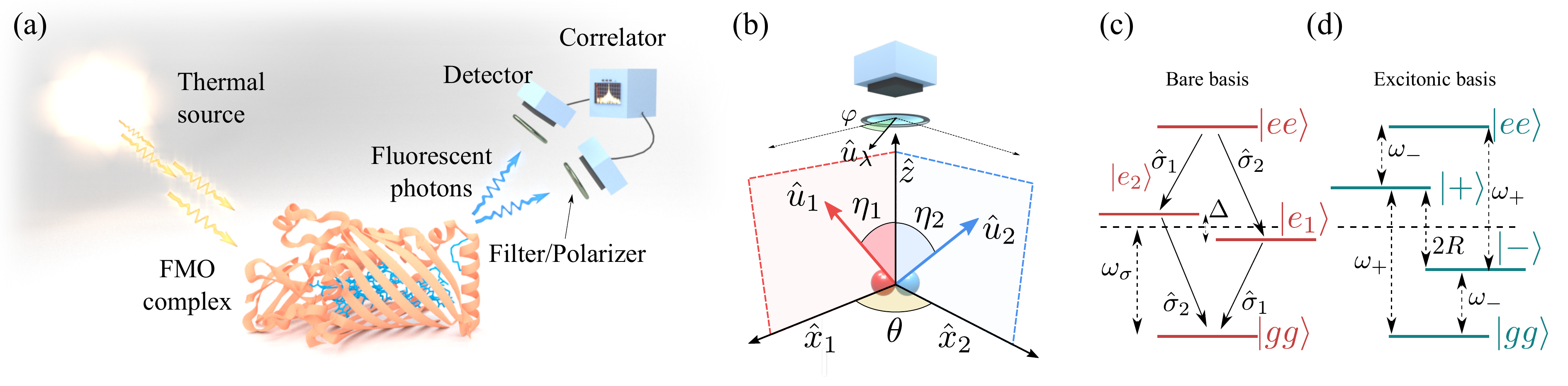}
\end{center}
\caption{(a) Sketch of the idea of photon-counting spectroscopy. Information about the state of a system $S$ (which might well be a stationary density matrix) is inferred through the statistics of photon-counting events. (b) Geometry of the dimer-detector system.  
(c) Bare basis  of our toy model~\eqref{eq:Hd}. (d) Excitonic basis which diagonalises the dipole Hamiltonian~\eqref{eq:Hd}. }
\label{fig:Setups}
\end{figure*}

\emph{Nonlinear laser spectroscopy vs. photon correlation measurements---}
On a microscopic level, nonlinear spectroscopy measures the nonlinear response functions of a sample, which can be connected to dipole correlation functions~\cite{Shaul_book}. For instance, standard techniques such as pump-probe measurements or photon echoes measure the third-order nonlinear susceptibility. Ultrashort pulses prepare a nonequilibrium state, whose evolution is monitored by a probe pulse after fixed time delay. This amounts to a measurement of response functions of the form $\sim \left\langle {d}_i (t_3) \left[ {d}_j (t_2), \left[ {d}_k (t_1), \left[ {d}_l (0),\varrho_0 \right] \right] \right]\right\rangle$, where $d_i$ denote components of the sample's dipole operator. 
Phase matching and control of the laser polarizations allows for additional selectivity, in order to only measure quantum pathways with the desired information~\cite{Ginsberg09, SchlauCohen11}, distinguish homogeneous from inhomogeneous broadening, or detect bath correlations from two-dimensional resonance lineshapes~\cite{Cho08}. Multi-color extensions of such measurements further enable the detection of correlations between electronic and vibrational dynamics~\cite{Lewis16}.

\begin{figure}[t!]
\begin{center}
\includegraphics[width=0.99\columnwidth]{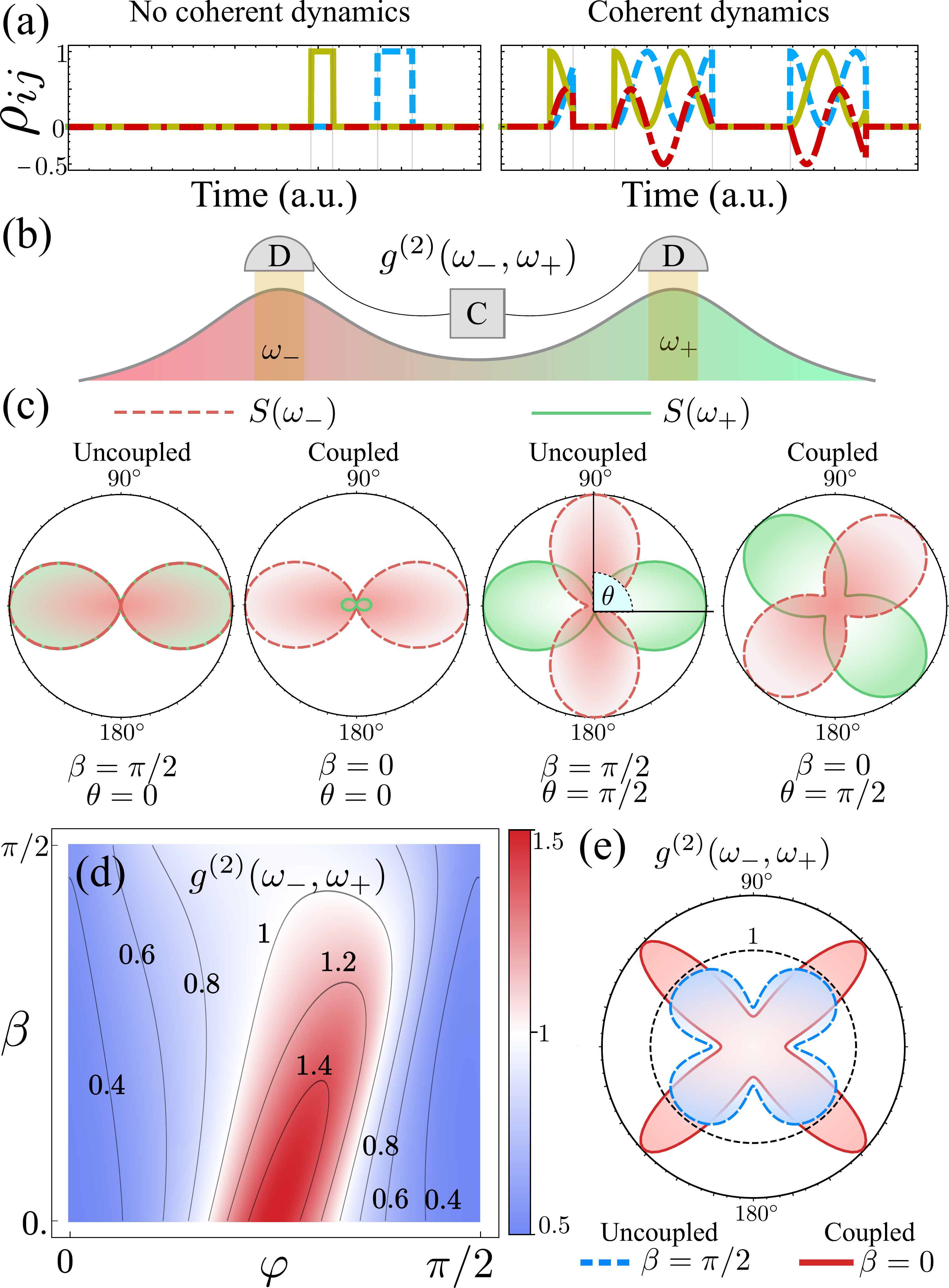}
\end{center}
\caption{(a) Monte Carlo trajectories depicting $\rho_{e_1 e_1}$ (solid, yellow), $\rho_{e_2 e_2}$ (dashed, blue), and $\rho_{e_1 e_2}$ (dot-dashed, red) during the process of absorbing and emitting several photons, with and without coherent couplings. For $P_1=P_2$, the oscillatory dynamics of the coherences average to zero steady-state coherence.  (b) Sketch of a frequency-resolved correlation setup measuring correlations between the two spectral peaks at $\omega_\pm $.
(c) Polarization measurements of the spectrum of emission at frequencies $\omega_\pm$ for different values of coherent coupling $J=R\cos\beta$ and inter-dipole angle $\theta$.  
 (d-e) Cross-correlation measurements at $\theta = \pi/2$ versus detection polarization for different values of $\beta$. Parameters: $R=2\gamma$, $P_1=P_2=0.1\gamma$, $\Gamma=\gamma$. }
\label{fig:spectrum}
\end{figure}

As we will see below, the conceptually simpler measurement of photon coincidences [see Fig.~\ref{fig:spectrum}(a)], a central quantity in quantum optics described by Glauber's second-order correlation function $g^{(2)}(t_1,t_2)$, contains similar information, as the fluorescence light can be directly translated into a function of the sample dipole operator. 
Measurements of bunching and anti-bunching statistics are widely employed to characterize quantum effects in continuously-driven cavity quantum electrodynamics systems~\cite{muller15a,hamsen17a}, and is recently gaining relevance in other fields, e.g. as a method to achieve super-resolution in biological imaging~\cite{tenne19a}. Its most general version meaures coincidences between photons detected at different energy-windows~\cite{delvalle12a,gonzaleztudela13a,sanchezmunoz14b,ulhaq12a,peiris15a,silva16a,Holdaway18},  described by the frequency-filtered Glauber's second-order correlation function~\cite{delvalle12a}:
\begin{equation}
  g^{(2)}(\omega_1,t_1;\omega_2,t_2)=\frac{\mean{:\mathcal{T}\big[\prod_{i=1}^2 \hat  E^{(-)}_{\omega_i,\Gamma}(t_i)\hat  E^{(+)} _{\omega_i,\Gamma}(t_i)\big]:}}{\prod_{i=1}^2\mean{\hat E^{(-)}_{\omega_i,\Gamma}(t_i)\hat E^{(+)}_{\omega_i,\Gamma}(t_i)}}\, ,\label{eq:g2-color}
\end{equation}
where $\mathcal T$ ($:$) refers to time (normal) ordering, and with $\hat E^{(\pm)}_{\omega_i,\Gamma}(t)$ the negative/positive frequency parts of the time-dependent electric field operator filtered at the frequency $\omega$ by a Lorentzian filter of linewidth $\Gamma$, $\hat E_{\omega,\Gamma}(t)=\frac{\Gamma}{2}\int_{0}^\infty \exp[-(i\omega+\Gamma/2)t']\hat E(t-t')\,\mathrm{d}t'$. Here, we will concerned with the simplest case of zero-delay statistics in the stationary regime, where the time-dependence in Eq.~\eqref{eq:g2-color} can be dropped, $g^{(2)}(\omega_1,\omega_2)\equiv \lim_{t\rightarrow \infty} g^{(2)}(\omega_1,t;\omega_2,t)$. Even in the zero-delay case, the frequency-filtered electric field involves an integral in time, therefore containing information not only about the stationary values of the density matrix $\rho$, but  also about the dynamics of the system, formally encoded in the Liouvillian superoperator $\mathcal L$ that generates the evolution of the density matrix, $\dot\rho = \mathcal L \rho$~\cite{sanchezmunoz2019a}. From Eq.~\eqref{eq:g2-color} and the definition of $\hat E_{\omega,\Gamma}(t)$, it is clear that photon correlation measurements too can be traced back to four-point correlation functions of the sample dipole operators, which is directly proportional to the radiated electric field (see below).
Therefore, as pointed out in Ref.~\cite{Dorfman18,Zhang18}, they can provide similar spectroscopic information as the measurement of the third-order nonlinear response using ultrafast laser, with the important advantage that, rather than using ultrashort pulses to initiate the dynamics, they can be performed in a steady state configuration.

\emph{Emission statistics from a single dimer---}
Now, we demonstrate the use of photon counting statistics to identify coherence in the simplest conceivable \emph{composite} quantum system: a dimer composed of two dipoles described in the two-level system (TLS) approximation [see Fig.~\ref{fig:Setups}(b)]. It forms the simplest toy model to describe the formation of excitonic states in molecular aggregates and their signatures in ultrafast spectroscopy. As a consequence, dimer systems have been widely studied in both theory~\cite{Kjellberg06, Pisliakov06, Chen10, Yuen-Zhou11, Bhattacharyya19} and experiments~\cite{Halpin14}, in order to better understand dynamic or spectroscopic features in more complex realistic models of light-harvesting complexes.

Each dipole has a dipolar moment operator $\hat{\mathbf{d}}_i=\mathbf{\mu}_i(\hat\sigma_i+\hat \sigma^\dagger_i)\mathbf{u}_i$, where $\hat\sigma_i$ is the lowering operator of the $i$-th TLS.
Considering the bare energies of the TLSs to be $\omega_\sigma$, and that any internal coupling rate is $g\ll \omega_\sigma$, the positive-frequency part of far-zone electric field operator radiated by the dimer is given by~\cite{scully_book02a} $\hat{\mathbf{E}}^{(+)}(\mathbf{r},t)= \mathbf{E_1}\hat\sigma_1(t-|\bm{r}|/c)+ \mathbf{E_2}\hat \sigma_2(t-|\bm{r}|/c)$,
with $\mathbf E_{i}\equiv E_r \mu_i \sin\eta_i \mathbf{x_i}$ and $E_r \equiv \omega_\sigma^2/4\pi\epsilon_0 c^2|\mathbf{r}|$.  This turns Eq.~\eqref{eq:g2-color} into a four-point dipole correlation function. The angles $\eta_i$ and the unit vector $\mathbf{x_i}$ are defined in Fig.~\ref{fig:Setups}(b).
In the following, we consider that the detectors measure a specific polarization $\mathbf u_\lambda$, 
yielding scalar fields $\hat E^{(+)}(\mathbf{r},t)= E_{1,\lambda} \hat\sigma_1(t-|\bm{r}|/c)+E_{2,\lambda}\hat \sigma_2(t-|\bm{r}|/c)$, where $E_{1(2),\lambda}\equiv \mathbf E_{1(2)}\cdot \mathbf u_\lambda$. Using the angle definitions shown in Fig.~\ref{fig:Setups}(b), and omitting from now on the subscript $\lambda$ for convenience, we can write this as
\begin{eqnarray}
 E_{1}&=& E_r \mu_1 \cos\varphi \sin\eta_1 \\
  E_{2}&=& E_r \mu_2 \cos(\theta-\varphi) \sin\eta_2 
 \label{eq:EBD_polarization}
\end{eqnarray}
Thus, $E_1$ and $E_2$ are determined by the relative orientation of the dipoles $\theta$, the polarizer angle $\varphi$, and the position of the detector. 
To simplify the discussion, we first consider photon counting statistics on a single detector, as sketched in Fig.~\ref{fig:Setups}(b). Thus, Glauber's zero-delay, second-order correlation function simply reads $g^{(2)}(0)=\langle \hat{E}^{(-)} \hat{E}^{(-)} \hat{E}^{(+)} \hat{E}^{(+)} \rangle /\langle \hat{E}^{(-)}\hat{E}^{(+)}\rangle^2$. Writing the correlators in terms of the elements of the density matrix as $\langle \sigma_1^\dagger \sigma_2^\dagger \sigma_1 \sigma_2 \rangle = \rho_{dd}$, $\langle \sigma_1^\dagger\sigma_1\rangle = \rho_{e_1e_1}+\rho_{dd}$, $\langle \sigma_2^\dagger\sigma_2\rangle = \rho_{e_2e_2}+\rho_{dd}$ and  $\langle \sigma_1^\dagger \sigma_2\rangle=\rho_{e_2e_1}$, we straightforwardly obtain
\begin{multline}
g^{(2)}(0) =4 (E_1 E_2)^2 \rho_{dd}/\left[E_1^2 \rho_{e_1e_1}+E_2^2\rho_{e_2e_2}\right.\\
\left.+(E_1^2+E_2^2)\rho_{dd}+2E_1E_2\Re{\rho_{e_2e_1}}\right]^2.
\label{eq:g2-site}
\end{multline}
\emph{Bounded photon fluctuations in uncoupled systems---}
Importantly, Eq.~\eqref{eq:g2-site} features the steady-state coherence $\rho_{e_2 e_1}$ in the denominator. This terms enters the equation through  $\langle \hat{E}^{(-)} \hat{E}^{(+)} \rangle$, which may suggest that a simple measurement of the intensity could suffice to determine the existence or absence of coherence. However, the intensity alone cannot distinguish coherence without an a-priori knowledge of the dipole strength of the sample.

On the other hand, a measurement of $g^{(2)}(0)$ can unambiguously verify the presence of excitonic coherence in the steady state. To prove this, we will now evaluate Eq.~\eqref{eq:g2-site} for a reference case in which no coherence exists. This reference case consists of two TLSs with no coherence, neither in the bare or the excitonic (energy) basis and no population imbalance, i.e. a state that fulfils the following three conditions: (I) $\rho_{e_1e_2}=\rho_{e_2e_1}=0$,  (II) $\rho_{e_1e_1}=\rho_{e_2e_2}\equiv  p$, and (III) $\rho_{dd} = p^2(1+p)^2$. 
The last equation comes from the assumption $\langle \sigma_1^\dagger\sigma_2^\dagger \sigma_1 \sigma_2\rangle = \langle\sigma_1^\dagger\sigma_1\rangle\langle\sigma_2^\dagger\sigma_2\rangle$. 
This is the expected situation when the dimer is excited by a light source with a bandwidth that is much larger than the detuning $\Delta$ between the two TLSs. For instance, for our discussion to remain valid for the excitation by sunlight, we require the detuning to be small enough for the variation of the light intensity with the wavelength to be negligible.
We can then show that photon correlations allow to identify states where these conditions are not fulfilled.  Applying these conditions in Eq.~\eqref{eq:g2-site}, we find $g^{(2)}(0) = 4 (E_1 E_2)^2/\left( E_1^2 +E_2^2 \right)^2 $, which we write in terms of the ratio $r = E_2/E_1$ as $\gdos = 4 r^2/(1+r^2)^2$. 
Its maximum value of $1$ is obtained at $r=1$. This is the expected result for two uncoupled TLS, in which the two emitters are independent of each other, which is an intuitive outcome in the absence of coherence. 
Hence, we find the criterion
\begin{equation}
\gdos>1
\label{eq:coherence-criterion-g2}
\end{equation}
  to identify coherence, i.e. the limit of two uncoupled emitters can only be surpassed in the presence of steady state coherence in the dimer.

\begin{figure}[b]
\begin{center}
\includegraphics[width=0.99\columnwidth]{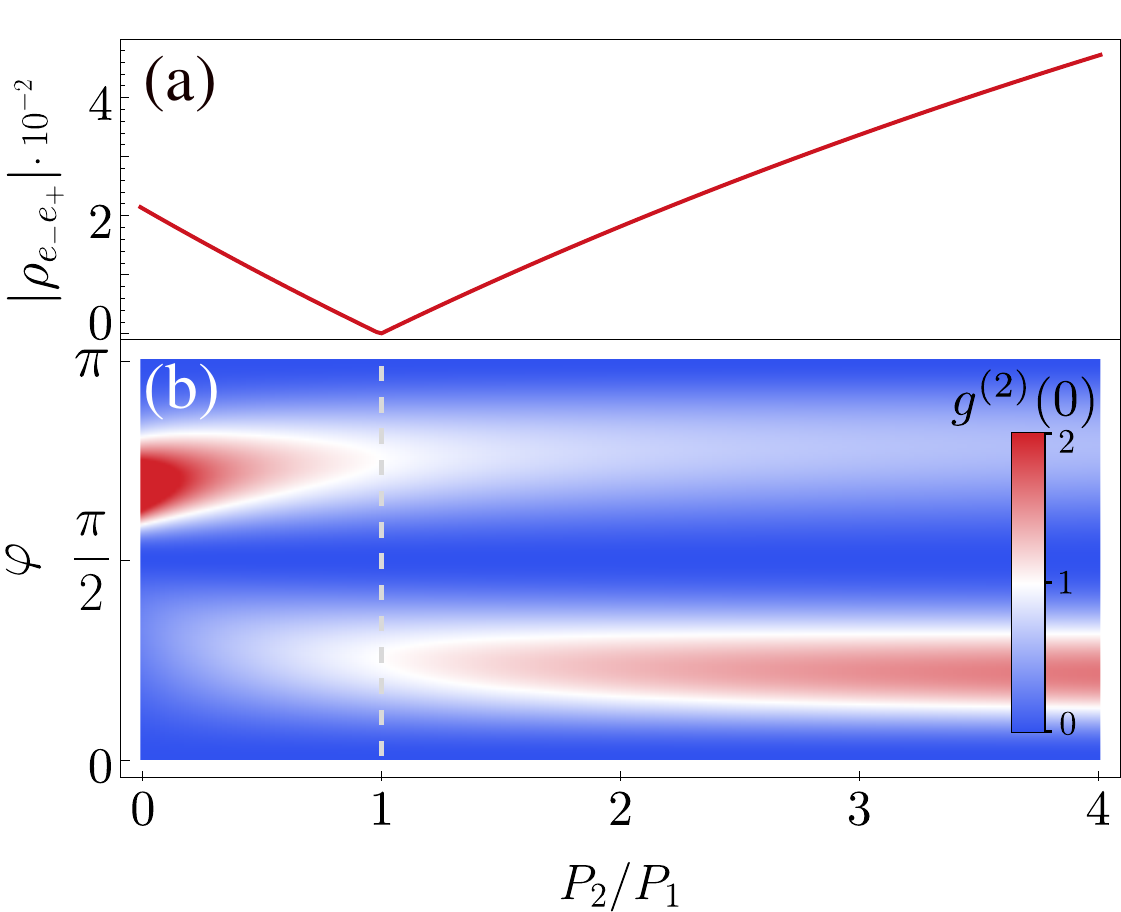}
\end{center}
\caption{(a) Coherence between excitonic states versus the pumping anisotropy $P_2/P_1$. When $P_1=P_2$, no excitonic coherence exists. (b) Dependence of $g^{(2)}(0)$ on the polarization angle $\varphi$ and the anisotropy $P_2$. Parameters: $R=4\gamma$, $\beta = \pi/3$, $J=2\gamma$, $\theta=\pi/2$, $P_1=0.1\gamma$. Coherence is evidenced by the existence of regions with $g_2^{(0)}>1$. }
\label{fig:g2-coherences}
\end{figure}

\emph{Specific model---} We now provide a particular example on how coherence can be detected by fulfilling this inequality. To this end, we introduce a  driven-dissipative dynamical model of the dimers in which we will tune the steady state coherence by introducing \textit{ad hoc} asymmetric incoherent pumping.

The model consists of two coupled TLSs with frequencies $\omega_\sigma\pm\Delta/2$, described by the dipole Hamiltonian
\begin{equation}
\hat H_d = (\omega_\sigma-\Delta/2)\hat\sigma_1^\dagger \hat\sigma_1+(\omega_\sigma+\Delta/2)\hat\sigma_2^\dagger \hat\sigma_2 + J(\hat\sigma_1^\dagger \hat\sigma_2 + \text{H.c.}),
\label{eq:Hd}
\end{equation}
where $J$ denotes coherent coupling between the TLSs.
The eigenstates for ${J=0}$ correspond to the bare or site basis, $\{|gg\rangle, |e_1\rangle, |e_2\rangle, |ee\rangle  \}$,  see Fig.~\ref{fig:Setups}(c). In the general case ${J\neq 0}$, the eigenstates in the one-excitation subspace are
$|+\rangle = c|e_1\rangle + s|e_2\rangle$ and $|-\rangle = s|e_1\rangle - c|e_2\rangle$, where $c=1/\sqrt{1+\xi^{-2}}$, $s=1/\sqrt{1+\xi^{2}}$, $\xi = J/(\Delta/2+R)$ and $R=\sqrt{J^2+(\Delta/2)^2}$, so that the corresponding energies are $\omega_\pm = \omega_\sigma \pm R$. We will consider the general case in which $R\neq 0$ and greater than the natural linewidths in the system, i.e. we will assume that one can observe two spectrally resolved peaks in the emission at the energies $\omega_\pm$. In order to keep the peak separation fixed, regardless of its possible origin, we parametrize $\Delta$ and $J$ by a mixing angle $\beta$ as $J=R\cos\beta$ and $\Delta/2=R\sin\beta$. Thus, the case $\beta=0$ corresponds to two peaks originating from two coupled, resonant dipoles, and $\beta=\pi/2$  to two uncoupled, detuned dipoles.

The driven-dissipative dynamics of the dimers in the presence of losses and incoherent driving is described by the master equation $\dot\rho = -i[H,\rho]+\sum_i\gamma\mathcal D_{\sigma_i}\{\rho\}/2
+\sum_i P_i \mathcal D_{\sigma_i^\dagger}\{\rho\}/2,$ with $\mathcal D_\mathcal{O}\{\rho\}\equiv 2\mathcal O \rho \mathcal O^\dagger - \mathcal O ^\dagger\mathcal O \rho-\rho\mathcal O^\dagger \mathcal O$, where $\gamma$ is the decay rate of both dimers and $P_i$ the incoherent pumping rate of the $i$-th dimer. Importantly, we allow for different incoherent excitation rates that can give rise to population imbalance. This is relevant since coherences in the excitonic basis are given by $\rho_{e_-e_+} = cs(\rho_{e_1e_1}-\rho_{e_2e_2})+s^2\rho_{e_2e_1}-c^2\rho_{e_1e_2}$. %
Thus, in order to have coherence in the excitonic basis we need coherence and/or population imbalance in the site basis, which we can enforce by setting $J\neq 0$ (so $c\neq 0$) and $P_1\neq P_2$. As shown in Fig.~\ref{fig:g2-coherences}(a), the pumping imbalance gives rise to steady state coherence between the excitonic states of the dimer which is directly proportional to the asymmetry of the pumps.
This allows us to illustrate the potential of $g^{(2)}(0)$ measurements to unveil coherences in the excitonic basis, as we demonstrate in Fig.~\ref{fig:g2-coherences}(b), where  $g^{(2)}(0)$ is plotted vs the anisotropy of the pumping and the polarizer angle of the detector setup. This plot reveals regions with $g^{(2)}(0)>1$ fulfilling the criterion Eq.~\eqref{eq:coherence-criterion-g2} and hence signalling the presence of steady state coherence in the system. Only when there is coherence in the steady state, the variation of the polarizer angle yields a value larger than unity for some angle that depends on the dimer parameters, and thus enables the direct detection of steady state coherence. This is the first main result of our paper.

\emph{Frequency-resolved measurements---}  
The previous analysis has been devoted to revealing coherence in the stationary density matrix of the system using photon correlation measurements. Now, we explore how frequency-resolved measurements [see Eq.~\eqref{eq:g2-color}] introduce a time resolution that allows to reveal transient coherent dynamics even when these average to zero in the steady state [see Fig.~\ref{fig:spectrum}(a)], allowing to distinguish this case from that of detuned, uncoupled emitters. In our model, we can describe coherent dynamics that yield zero stationary coherences by setting $J\neq 0$, $P_1=P_2$, which is the case we consider from now on. 

From now on, we will be concerned with the light emitted at the two possible transition energies between eigenstates, $\omega_\pm$, see Fig.~\ref{fig:spectrum}(b). 
Figure~\ref{fig:spectrum}(c) shows the polarization profile of the light emitted at these two frequencies, given by $S(\omega_\pm)=\lim_{t\rightarrow \infty}\langle \hat{E}^{(-)}_{\omega_\pm,\Gamma}(t) \hat{E}^{(+)}_{\omega_\pm,\Gamma}(t) \rangle$, for several combinations of $\beta$ and $\theta$. If we consider that most of the emission will come from the single-excitation subspace, so that $S(\omega_\pm) \propto  |\langle gg|\hat{E}^{(+)}|\pm\rangle |^2 $, we expect
\begin{equation}
S(\omega_{\pm}) \propto E_{1/2}^2 c^2 + E_{2/1}^2 s^2 \pm 2cs E_1 E_2.
\label{eq:S_omega_-}
\end{equation}
These equations can qualitatively explain the features in Fig.~\ref{fig:spectrum}(c). Namely, for $\theta = 0$, the two dipoles are aligned, 
so both peaks exhibit the same dependence with the polarizer angle $\varphi$. For $\theta = \pi/2$, the two are orthogonal, $E_1\propto \cos\varphi$, $E_2 \propto \sin\varphi$. A frequency-resolved polarization measurement therefore provides valuable information about the internal structure of the dimer, since the angle between the two lobes gives an estimate of $\theta$. However, the information obtained about the presence of coherent dynamics  is limited: only in the case of aligned dipoles and $\beta=\pi/2$ the presence of two peaks with same $\varphi$-dependence unambiguously signifies the absence of coherent coupling.
All the other cases shown in Fig.~\ref{fig:spectrum}(c), are ambiguous, i.e., one cannot tell whether the peaks are created by uncoupled or coupled emitters. 

Coherent couplings can, however, be unveiled by correlation measurements in the frequency domain. Without loss of generality, we focus on the case $\theta=\pi/2$. We analyse cross-correlations between the two emission peaks [see Fig.~\ref{fig:spectrum}(b)], since any possible two-photon de-excitation process from $|ee\rangle$ will involve photons of the two frequencies $\omega_\pm$ [see Fig.~\ref{fig:Setups}(c)]. As we show in Figs.~\ref{fig:spectrum}(d-e), measuring this cross-peak correlation for different values of $\varphi$ reveals bunching features [$g^{(2)}(\omega_+,\omega_-)>1$] around $\varphi = \pi/4$  only in the presence of coherent coupling, even if the steady-state coherence is zero in all cases. To qualitatively understand this observation, we note that, at $\varphi=\pi/4$, coherent coupling yields a destructive interference of the emission at $\omega_-$ originating from the first-excitation subspace, i.e. $S(\omega_-)=0$ in Eq.~\eqref{eq:S_omega_-}. Consequently, any emission detected at $\omega_-$ must originate from a two-photon cascade from the doubly-excited state $|ee\rangle$ to state $|+\rangle$, and will therefore show strong correlations with subsequently emitted photons from $|+\rangle$ to $|gg\rangle$ at frequency $\omega_+$. In the absence of coupling, the two peaks must originate from two independent, detuned emitters, and one recovers the expected result for uncorrelated emission, $g^{(2)}(\omega_+,\omega_-)\approx 1$~\footnote{Deviations from this result are due to finite-filter effects.}. Note the qualitative difference between coupled and uncoupled cases in Fig.~\ref{fig:spectrum}(e) as compared to the plots in panel (c). 
\\
\emph{Conclusions---} We have established the potential of steady-state photon correlation measurements as a novel tool for molecular spectroscopy. The analysis of the emission statistics of a dimer model revealed that it is possible to detect the presence of environment-induced steady state coherence in the system. In frequency-resolved measurements, we observe that, akin to cross-peaks in two-dimensional laser spectroscopy~\cite{Cho08}, the bunching of photons at two different frequencies indicates coherent coupling between the dipoles. The strength of the coherent coupling can be read off the bunching ratio, and provides direct access to the bare emitter frequencies, whose accurate determination is of critical importance to relate structural and optical properties in molecular aggregates \cite{Adolphs07, Cheng09,  Milder10}. Our results clearly demonstrate that photon correlations can provide a new valuable tool for experimentalists and open new avenues in the field of quantum spectroscopy~\cite{Dorfman16, AccChemRes}.

\paragraph*{Acknowledgements} 
C.S.M. is funded by the Marie Sklodowska-Curie Fellowship QUSON (Project  No. 752180). F. S. acknowledges funding from the European Research Council under the European Union's Seventh Framework Programme (FP7/2007-2013) Grant Agreement No. 319286 Q-MAC.

\let\oldaddcontentsline\addcontentsline
\renewcommand{\addcontentsline}[3]{}
\bibliographystyle{mybibstyle}
\bibliography{bibliography_photons,sci-url,books}

\let\addcontentsline\oldaddcontentsline

\clearpage
\onecolumngrid
\appendix

\begin{center}
{\bf \large Supplementary Material}
\end{center}

\renewcommand{\theequation}{S\arabic{equation}}

\renewcommand{\thefigure}{S\arabic{figure}} 
\setcounter{figure}{0} 
\setcounter{equation}{0}   

\tableofcontents

\section*{Analytical expressions for steady state density matrix elements}
The master equation analysed in the main text 
\begin{equation}
\dot\rho = -i[H,\rho]+\sum_i\gamma\mathcal D_{\sigma_i}\{\rho\}/2
+\sum_i P_i \mathcal D_{\sigma_i^\dagger}\{\rho\}/2
\end{equation}
yields the following stationary expectation values:
\begin{subequations}
\begin{align}
&\langle \sigma_1^\dagger \sigma_1 \rangle =\frac{4J^2(\tilde P_1 + \tilde P_2)(P_1+P_2)+P_1 \tilde P_2 \chi}{4J^2(\tilde P_1+\tilde P_2)^2+\tilde P_1 \tilde P_2 \chi}, \\
&\langle \sigma_2^\dagger \sigma_2 \rangle =\frac{4J^2(\tilde P_1 + \tilde P_2)(P_1+P_2)+\tilde P_1 P_2 \chi}{4J^2(\tilde P_1+\tilde P_2)^2+\tilde P_1 \tilde P_2 \chi}, \\
&\langle\sigma_1^\dagger \sigma_2\rangle = \frac{-2iJ\gamma(\tilde P_1-\tilde P_2)[\tilde P_1+\tilde P_2+ 2i\Delta]}{4J^2(\tilde P_1+\tilde P_2)^2+\tilde P_1 \tilde P_2 \chi},
\label{eq:coherence-ss}
\\
&\langle \sigma_1^\dagger \sigma_2^\dagger \sigma_1\sigma_2\rangle = \frac{4J^2(P_1+P_2)^2+P_1 P_2 \chi}{4J^2(\tilde P_1+\tilde P_2)^2+\tilde P_1 \tilde P_2 \chi},
\end{align}
\end{subequations}
where $\tilde P_i \equiv P_i+\gamma$ and $\chi \equiv (\tilde P_1+\tilde P_2)^2+4\Delta^2$. It is important to notice that coherences $\rho_{e_1 e_2}$, given by the correlator $\langle \sigma_1^\dagger \sigma_2\rangle$, are only different from zero provided $J\neq 0$ and $P_1\neq P_2$. The later condition is also a requirement to obtain steady-state population imbalance on the bare basis, which is given by:
\begin{equation}
\rho_{e_1 e_1}-\rho_{e_2 e_2} =  \frac{(P_1-P_2)\gamma \chi}{4J^2(\tilde P_1+\tilde P_2)^2+\tilde P_1 \tilde P_2 \chi}.
\end{equation}

As we discuss in the main text, excitonic coherence is given by
\begin{equation}
\rho_{e_-e_+} = cs(\rho_{e_1e_1}-\rho_{e_2e_2})+s^2\rho_{e_2e_1}-c^2\rho_{e_1e_2},
\end{equation}
which means that driving imbalance $P_1\neq P_2$ is also a requisite for excitonic coherence, which can therefore by tuned by introducing \emph{ad hoc} asymmetric incoherent driving.

\section*{Detection of coherence using $g^{(2)}(0)$: Dependence on system parameters}

\begin{figure*}[t]
\begin{center}
\includegraphics[width=0.8\columnwidth]{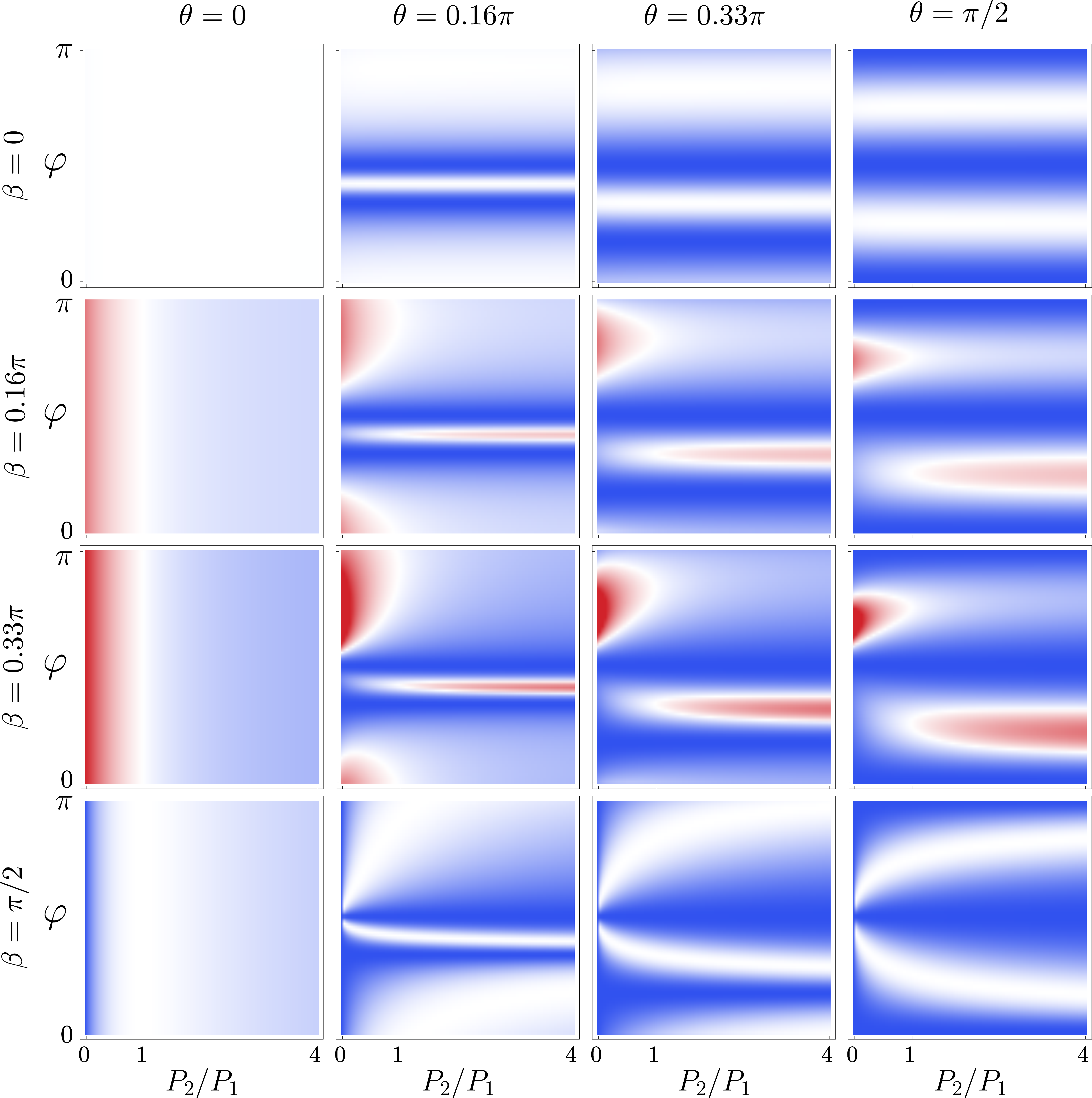}
\end{center}
\caption{ Dependence of $g^2 (0)$ on the polarization angle and the anisotropy $P_2 / P_1$, as in Fig.~\ref{fig:g2-coherences}(b). The two free parameters of the dimer system, the mixing angle $\beta$ and the angle between the dipoles $\theta$ are varied.}
\label{fig:g2-parameters}
\end{figure*}

In Fig.~\ref{fig:g2-parameters}, we extend the simulations in Fig.~\ref{fig:g2-coherences} in the main text to different combinations of the angle between bare dipoles $\theta$ and mixing angle $\beta$. We find that $\theta$ does not impede our ability to detect steady state coherence, i.e. if bunching $g^{(2)}(0)>1$ signalling coherence is measured for a given dipole-dipole angle $\theta$, it would also be measured for any other angle. Coherence is evidenced by the criterion $g^{(2)}(0)>1$ except in the limit case $\beta = 0$.

\section*{Frequency-resolved photon correlation measurements}

In the main text, we have provided an intuitive explanation for our findings on the frequency-resolved correlations between peaks. Beyond this explanation, we can also qualitatively describe our findings by computing transition matrix elements as we did for the spectrum peaks $S(\omega_\pm)$. To do this, we consider, at zero time delay, that the quantity $G^{(2)}(\omega_+,\omega_-)=\langle E^{(-)}_{\omega_+,\Gamma}
E^{(-)}_{\omega_-,\Gamma} E^{(+)}_{\omega_+,\Gamma} E^{(+)}_{\omega_-,\Gamma} \rangle $ in the numerator of Eq.~(\ref{eq:g2-color}) is given by
\begin{equation}
G^{(2)}(\omega_+,\omega_-)\propto|\sum_{j=+,-}\langle gg|\hat{E}^{(+)}|j\rangle\langle j|\hat{E}^{(+)}|ee\rangle|^2  \propto |E_1 E_2|^2 \propto \cos^2\varphi \cos^2(\theta-\varphi).
\end{equation}
To see how this qualitatively explains the dependence of $g^{(2)}(\omega_+,\omega_-)$ on the polarisation angle, $\varphi$, and the mixing angle, $\beta$, let us consider the two following limiting cases:
\paragraph{$\beta = 0$ (strongly-coupled)}
For $\beta = 0$, we have $J=R$, and $c=s=1/\sqrt 2$, so $S(\omega_+)S(\omega_-)\propto \left[ \cos^2\varphi - \cos^2(\theta-\varphi)\right]^2 $, giving 
\begin{equation}
g^{(2)}(\omega_1,\omega_2)\propto \frac{\cos^2\varphi \cos^2(\theta-\varphi)}{\left[ \cos^2\varphi - \cos^2(\theta-\varphi)\right]^2}.
\end{equation}
For $\theta = \pi/2$, this is equal to
\begin{equation}
g^{(2)}(\omega_1,\omega_2)\propto \frac{\cos^2\varphi \sin^2\varphi}{\left( \cos^2\varphi - \sin^2\varphi\right)^2}.
\end{equation}

\paragraph{$\beta=\pi/2$ (uncoupled)}
For $\beta = \pi/2$, $J=0$, $c=0$, $s=1$, giving $S(\omega_+)S(\omega_-)= E_1^2 E_2^2$, and $g^{(2)}(\omega_1,\omega_2)\propto 1 $. 
\\

\begin{figure}[t!]
\begin{center}
\includegraphics[width=0.5\textwidth]{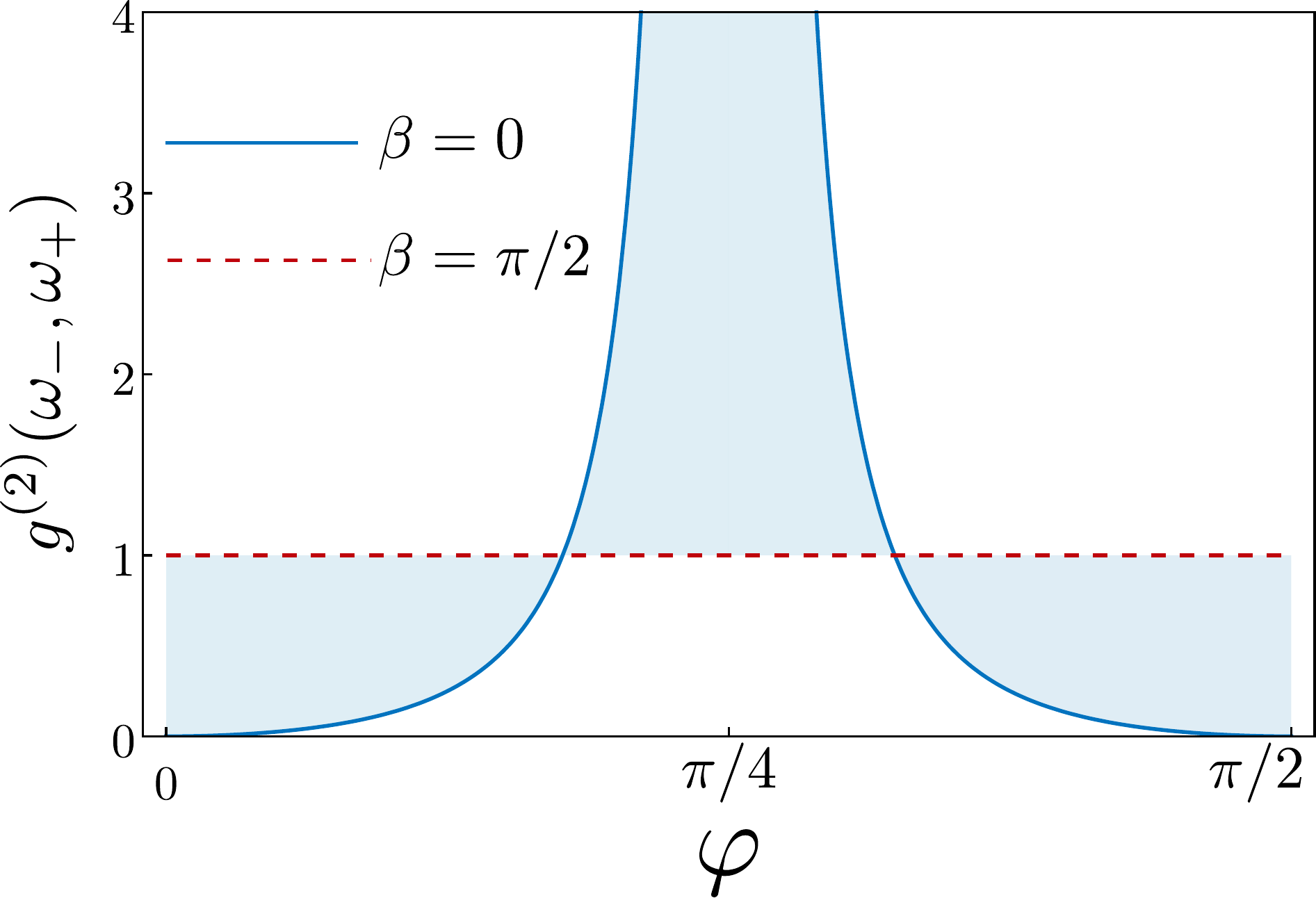}
\end{center}
\caption{Analytical estimation of the frequency-resolved correlation function between the two emission peaks, $g^{(2)}(\omega_-,\omega_+)$ for the two limiting cases of strongly-coupled ($\beta = 0$) and uncoupled ($\beta=\pi/2$) emitters. }
\label{fig:g2-analytical}
\end{figure}

These two possible expressions are plotted in Fig.~\ref{fig:g2-analytical}, and qualitatively replicate the characteristic feature observed in our exact, numerical calculations shown in in Fig.~\ref{fig:spectrum}(d), namely the emergence of bunching in the presence of coherent dynamics. The expressions obtained here only qualitatively match the exact result due to the contributions to the spectrum $S(\omega_\pm)$ from the doubly excited state $|ee\rangle$, not considered in these analytical estimations, and to the finite filter linewidth $\Gamma$.

\end{document}